# Viscoelastic properties of wood across the grain measured under water-saturated conditions up to 135°C: evidence of thermal degradation

[1]Vincent Placet, [2]Joëlle Passard, [2]Patrick Perré

[1] CNRS UMR 6174 FEMTO-ST Institute, Department of Applied Mechanics, 24 Chemin de l'Epitaphe. F-25000 Besançon.
[2] UMR1093 Etudes et Recherche sur le Matériau Bois, INRA, ENGREF, 14, rue Girardet. F-54042 Nancy.

**Abstract**

In this paper, the viscoelastic properties of wood under water-saturated conditions are investigated from 10°C to 135°C using the WAVE$^{T*}$ apparatus. Experiments were performed via harmonic tests at two frequencies (0.1 Hz and 1 Hz) for several hours. Four species of wood were tested in the radial and tangential material directions: oak (*Quercus sessiliflora*), beech (*Fagus sylvatica*), spruce (*Picea abies*) and fir (*Abies pectinata*).

When the treatment is applied for several hours, a reduction of the wood rigidity is significant from temperature values as low as 80-90°C, and increases rapidly with the temperature level. The storage modulus of oak wood is divided by a factor two after three hours of exposure at 135°C. This marked reduction in rigidity is attributed to the hydrolysis of hemicelluloses.

The softening temperature of wood is also noticeably affected by hygrothermal treatment. After three short successive treatments up to 135°C, the softening temperature of oak shifted from 79°C to 103°C, at a frequency of 1 Hz. This reduction in mobility of wood polymers is consistent with the condensation of lignins observed by many authors at this temperature level.

In the same conditions, fir exhibited a softening temperature decreasing of about 4°C. In any case, the internal friction clearly raises.

*WAVE$^T$ : Environmental Vibration Analyser for Wood.*

V. Placet, Department of Applied Mechanics, University of Franche-Comté, F-25000 Besançon, vincent.placet@univ-fcomte.fr

# 1. Introduction

Wood is a natural composite material characterised by a complex biochemical organisation, a polymer blend made up of three main polymers, i.e. cellulose, hemicellulose and lignin. Some of those exhibit a linear arrangement, namely cellulose and hemicelluloses, while, lignins are branched and constitute a crosslinked network. Moreover, hemicelluloses and lignin are completely amorphous while cellulose is partly crystalline. The rheological behaviour of wood results from the intricate combination of each polymer component response. Contrary to classical homogeneous polymers, the concept of glass transition temperature does not apply to wood. Instead, we are talking of a softening temperature, a temperature range in which the rheological properties vary rapidly.

In addition, wood is an hygroscopic material which contains bound water. The viscoelastic behaviour of wet wood is essential in many sectors of the wood industry. It particularly concerns wood processing operations such as drying, wood forming, thermal treatment, veneer and manufacturing based on wooden materials like panel pressing, wood gluing and so on…

From the rheological point of view, water is a plasticizing agent for wood. The replacement of hydrogen bonding within the amorphous components by water-carbohydrates links enhances the flexibility of the polymer network. As a consequence, water-saturated wood depicts a lower rigidity and a lower softening temperature than dry wood [1, 2]. In water-saturated conditions, the properties of wood reflect to a large extent the properties of the wet lignin. Indeed, Salmén [3], Kelley *et al.* [1], showed that the main wood softening, appearing between 50°C an 100°C, corresponds to the glass transition of *in situ* lignin. Secondary relaxations corresponding to the softening of the other carbohydrates (amorphous cellulose and hemicelluloses) appear at lower temperature [4]. But as the wood dries, the lignin softening effect is reduced and the hemicellulose softening is more important [4, 5]. Beyond the water content level, the moisture variations (i.e. the mechano-sorption) also affect the mechanical creep [6, 7]. To this day, the origin of this phenomenon is only understood to some extent by the wood community. Several models attempt to describe this complex phenomenon [8-10]. Moreover, more recent work indicates potential interactions between viscoelastic and mechano-sorptive creep [11].

So, the viscoelastic properties and particularly the softening temperature of wood, is strongly affected by moisture content and moisture variations and also by temperature. Actually, as for most polymers, the temperature level dramatically affects the viscoelastic properties. Heat allows inter-molecular linking to be broken, allows for molecular mobility and therefore provides more flexibility to the macromolecular network. Crawling movements are also possible between macromolecular chains, and the wood softens.

However, one may wonder if one part of the apparent thermal activation may result from thermal degradation. The degradation mechanisms of wood also depend on temperature and moisture levels. In the case of dry wood, thermal degradation is significant only above 200°C, but this threshold value decreases down to around 100°C when soaked samples remain hot for several hours [12, 13]. So, the question is to know by what extend thermal degradation alters the viscoelastic properties determined on water-saturated wood near and above 100°C.

Thermal softening of wet wood has previously been study above 100°C by many researchers namely by quasi-static methods [14, 15, 16, 17]. In comparison with quasi-static tests, harmonic tests represent a more efficient method to characterise viscoelastic properties and glass transition of materials. By allowing the mechanical time to be different from the experiment time, harmonic tests are capable of distinguishing the viscoelastic behaviour from other effects such as ageing or recovery of growth stress. Certainly due to the difficulty to perform this type of test above 100°C in water-saturated conditions, data collected by this method in such severe hygro-thermal conditions are really rare in the literature. Indeed, only one reference was found [3].

The subject of this paper is to investigate the mechanical response of wood across the grain affected by severe hygro-thermal treatment, in water saturated conditions below 135°C. The measurements of the viscoelastic properties were performed with a dynamic mechanical analyser specially conceived in our laboratory for wood, the WAVE$^T$ (Environmental Vibration Analyser for Wood).

## 2. Material and method

Several wood species were tested in this work: oak (*Quercus sessiliflora*), beech (*Fagus sylvatica*), spruce (*Picea abies*) and fir (*Abies pedtinata*). There were harvested in the eastern part of France, in settlements belonging to ENGREF (the Forest School of Nancy). Samples were cut in the heartwood part of green logs along the two material directions across the grain, i.e. radial and tangential directions. The cross section of the samples is $6 \times 10$ mm² and its length 100 mm. Care was taken to select normal wood, that is to say mature wood without any reaction wood or singularity.

Measurements were conducted with the WAVE$^T$ device, a custom apparatus developed in our laboratory and perfectly adapted to wood features, namely its hygroscopicity and anisotorpy. A detailed description of this apparatus is presented in a related paper, Placet *et al*. [18] and in Placet [19], and many results collected for green wood using the device below 100°C can be found in Placet *et al*.[20].

The WAVE$^T$ assumes the dynamic mechanical analysis of samples in flexion (single cantilever beam) in water-saturated conditions. In its first version, tests were led in water at atmospheric pressure. A second version has been especially developing to perform tests under overpressure, at a boiling point of water up to 140°C (Fig. 1). The entire system, composed of the sample holder, the actuator and the sensors, is placed in a pressurised conditioning room. This specific chamber allows the sample to be immersed in water at temperature level reaching 140°C while maintaining mild conditions around the electronic apparatus (temperature near 20°C and relative humidity around 10%). Non-airtight communicating areas exist between the two compartments to carry out the sample loading and the force and deflection measurements without friction. This conditioning system is the subject of a patent application [21].

In dynamic mechanical analysis, the sample is subjected to sinusoidal load with various frequencies. The temperature at which a maximum of energy is dissipated denotes the largest movements of the polymer chains. This temperature is the glass transition temperature, noted Tg. It indicates that molecular movements have the same relaxation time as the applied frequency. As explained before, for heterogeneous polymer such as wood, we don't talk of a glass transition temperature but of a softening temperature. Indeed, because the wood behaviour results from the combination of several polymers, the transition zone is less marked than for pure polymers.

Results of dynamic mechanical analysis are commonly presented as storage modulus (E'), loss modulus (E'') and loss factor (tan$\delta$ = E''/E'). The storage modulus describes the capacity of material to support a load, and so represents the elastic part of the sample. The loss modulus is the viscous response of the sample and is proportional to the dissipated energy. The loss factor characterises the damping capacity of the material. When scanning in temperature, the softening temperature is characterized by a decrease in the storage modulus and a peak of E'' and tan$\delta$. In this work, taking into account that only the tan$\delta$ peak is always observable in the softening area, we identify the softening temperature using the peak value of the loss factor. Although a more rigorous analysis should apply to define the softnening temperature, many authors used the same rule in the case of wood [1, 3, 4].

The WAVE$^T$ is able to determine the viscoelastic properties (stiffness and damping properties) of water-saturated samples between 5°C to 140°C, at frequencies $5.10^{-3}$ to 10 Hz and at stress levels 0.01 to 4 MPa. In this forced non-resonance technique, the samples are immersed in a temperature-controlled water bath throughout the test, at a pressure increasing up to 5 bar. Specimens are tested in a bending configuration based on a rigorous single cantilever mode: a specific articulated clamp ensures that a pure vertical force is applied without any momentum to the sample. Using the strength of materials relations [22], we can express the deflection relationship according to stress and sample geometry in this deformation mode (Eq. 1).

$$H = H_M + H_T$$

$$H = \frac{F l_0^2 (6L - 2l_0)}{bh^3 E} + \alpha \frac{F l_0}{bh} \frac{1}{G} \quad (1)$$

$H_M$ Deflection due to the bending moment

$H_T$ Deflection due to the shear stress

F Load (N)

L Distance of the load application point from the support (m)

$l_0$ Distance of the deflection measuring point from the support (m)

b width of the beam (m)

h thickness of the beam (m)

E Modulus of elasticity (Pa)

$\alpha$ coefficient which depends on the shape of the specimen (3/2)

G Coulomb's modulus

Passard and Perré [16, 17] showed for the transverse material directions, the effect of shear stress can be negligible. In consequences, equation 1 is simplified as following:

$$H = \frac{F l_0^2 (6L - 2l_0)}{bh^3 E} \quad (2)$$

Moreover, contrary to most commercial instruments, the force applied to the sample is not calculated from knowledge of the input signal to the driver but actually measured by a miniature load cell placed between the sample and the driver. As additional advantage, this load cell allows the applied force to be controlled by a closed-loop regulation, especially relevant for very low frequencies. Finally, in order to avoid any disturbance at the clamp area, the deflection measurement is dissociated from the other part of the system.

Besides, great care was taken to collect, treat, and analyse raw data. The phase difference, a key parameter, is identified by an inverse method over several periods. In addition, a comprehensive model of the whole device is implemented in the control software to account for the frame rigidity and the inertia of the moving part and correct their effects in the measured parameters.

The sinusoidal vibrations are performed with a zero mean stress in order to avoid a deflection due to additional creep. In all cases, samples are tested within the linear viscoelastic range. Notice that the strain value varies between 0.01% and 0.1%. The viscoelastic parameters are determined taking into consideration of the macroscopic dimensions of the wood specimens at 20°C.

Figure 2 exhibits typical results obtained using this device. The softening temperature is obvious on these curves and one has to notice that the unique ability of the device to work above 100°C allowed the tan$\delta$ peak to be clearly determined even in the case of softwood.

Two types of tests were performed to characterise the mechanical response of wood submitted to severe hygro-thermal treatments. First, a set of samples in radial and tangential directions has been tested at different plateau temperatures (100°C, 110°C, 135°C) for two frequencies (1 and 0.1 Hz) during 18 to 26 hours according to samples. Viscoelastic properties are measured every four minutes. The heating rate used to reach the plateau temperature is 1.25°C.min$^{-1}$. For each species and material direction, several samples have been analysed to check the measurement repeatability.

The second series of tests consisted in submitting green samples to successive rising and falling temperature stages. Temperature scans utilised a heating and a cooling rate of 0.35°C.min$^{-1}$ between 10°C and 135°C at a frequency of 1 Hz. Viscoelastic properties were measured every 5°C. These temperature cycles are repeated three times and allowed the degradation occurring to high temperature level to be characterised.

It is well known that material history and particularly water content variations affects the viscoelastic properties of wood. Therefore, samples were always prepared from a green log, immediately soaked in cold water in an airtight box and tested within a few days. We also prevented any drying to occur during the sampling.

## 3. Results and discussion

Measurements performed with the first version of the WAVE$^T$, in water-saturated conditions below 100°C revealed the existence of some degradation in wood for tests led above 80-90°C during several hours. Figure 3 depicts the evolution of the storage modulus of beech samples submitted to hygro-thermal treatment at three soft plateau temperatures (64°C, 78°C, 93°C). After 8 hours, the decrease of the storage modulus is close to 3% at 62°C, but attains a value of about 15% at 93°C. By decoupling the mechanical characteristic time (test frequency) from the degradation characteristic time (duration of the plateau at the selected temperature) these results show that the temperature level acts as an activator of the viscoelastic properties but also as a degradation factor, namely with soaked samples. This observation highlights the interest to perform similar experiments above 100°C.

The mechanical behaviour of wood versus time at different plateau temperatures above 100°C gives unexpected results. Figure 4 and 5 depict the evolution of the storage modulus for spruce and oak samples in radial and tangential directions. For oak, below 120°C, the rigidity drops with the experiment duration. This phenomenon is attributed to the degradation of the hemicelluloses by hydrolysis [12, 23]. Actually, the hemicelluloses react more sensitively than cellulose during heating. For a temperature of 110°C, two degradation time constants may be distinguished: a rapid decrease during the first four hours and slower degradation, where the decrease in time is almost linear, during the following hours. At 135°C, the results are somehow surprising. After a consistent decreasing of the storage modulus during the first two hours of thermal treatment, the material rigidity tends to increase during the following hours. After a reduction by 50% after 3 hours, the storage modulus returns to its initial value after 24 hours. According to the results of Funaoka *et al*. [12], this is certainly due to the condensation of lignins, that is to say by the increasing of the strong links (or condensed links) between lignin units. It is well known that condensation is accelerated by the presence of water and by the decrease in pH, which here is due to the hydrolysis of hemicelluloses. We can also imagine that the macromolecular reorganisations due to the massive removal of a component of weak rigidity as hemicelluloses may led to a stronger structure.

Small differences are observed between the radial and the tangential directions. The decreasing followed by the increasing in the storage modulus at 135°C are more accentuate in the tangential direction during the first ten hours. Table 2 gives the values of the storage modulus at different time as a function of the material direction.

The ratio $E'_R/E'_T$ is just slightly higher than the unit for oak at these temperature levels compare to the ratio value obtained at 20°C (1.4).

Figure 6 shows the dramatically high shrinkage level of tested samples after drying with moderate conditions at room temperature. This is certainly due to collapse provoked by the loss of rigidity of the cell wall. As stated by Passard & Perré [13], collapse is a good indicator of both thermal softening and thermal degradation. On one hand, the thermal activation of the viscoelastic properties decreases the storage modulus of the cell wall; this latter is no more capable of supporting the capillary pressure inside the lumens. On the other hand, the mechanical resistance of the cell wall is weakening because of the degradation of the hemicelluloses. ESEM photographs (Fig. 6) confirm the shape of collapsed tissues, that flattened the vessels in the tangential direction (vessels are almost circular in native wood). At higher magnification (right side), it seems that the fibre lumens disappeared and fibre cell wall are indistinct, as if the wall layers had melted.

The kinetics of the spruce degradation differs from that of oak. This is not surprising if we keep in mind that the structural organisation of softwoods and particularly their biochemical organisation is by several extends different than that of hardwoods. From the rheological point of view, the important cross-linking of the sprucewood lignin units (no S-units in softwood, Fig. 7) builds up a rigid network that restricts the chain mobility. Molecular mobility is unlikely to occur unless numerous bonds are broken simultaneously. Then, the activation energy required to soften sprucewood is high. This explains that hardwood lignins soften at a lower temperature than softwood lignins (Fig. 2).

Compared to oak, Table 3 and Figure 5 show that the decreasing of the storage modulus attributed to the thermolysis of hemicelluloses is less important for spruce. In addition, the increasing due to the condensation of lignin appears at a lower temperature level (110°C). In tangential direction, at 110°C, after a decreasing in rigidity of about 20%, the value starts to increase in time to finally exceed the initial value (Table 3). The ratio $E'_R/E'_T$ remained quite high at this temperature level. This could be explained by the fact the ratio at 20°C (near 2.5) is much higher than the ratio measured for oak (1.4).

No fundamental difference has to be noted between the tests performed at 1 and 0.1 Hz. The trends are the same, with just a consistent effect of the frequency on the storage modulus.

Thus, being aware of the thermal degradation of wood, it is possible to wonder about the influence of the thermal degradation on the measurements of the viscoelastic properties versus temperature. Actually, even for measurements between 10°C and 135°C at 1 Hz (hence without additional time required by very low frequencies), the effect of thermal degradation is noticeable (Fig. 8). On this Cole-Cole diagram, it becomes obvious that measurements performed at high temperature include one part of thermal degradation: the points determined above 100°C are outside the arc of ellipse characteristic of the *in situ* lignin transition.

Successive temperature cycles allow differences in wood material responses due to hygro-thermal treatment to be characterised. The first rise in temperature reveals the *in situ* lignin transition, between 70 and 105°C, by the tanδ peak accompanied by a drop in the storage modulus appears (Fig. 9 and 10). As explained sooner and because of theirs native lignin structure, the softening of untreated samples is detected at higher temperature for softwoods (fir and spruce) than for hardwood (oak). After successive cycles, the effect of the severe conditions (elevated temperature and presence of liquid water) becomes evident, both for oak and for spruce.

For oak in radial direction, the initial softening temperature of 79°C is translated to 103°C, and the storage modulus is divided by two after three temperature cycles. Internal friction noticeably raises after the first exposure to high temperature; the tanδ maximum goes from 0.126 to 0.163. The decreasing of the storage modulus is lower as the cycles accumulate. At 20°C, the initial storage modulus is equal to 945 MPa. After the first cycle, the storage modulus at 20°C decreased by about 33%, after the second by 24% and after the third by 10%.

These results are consistent with the hypothesis of the hydrolysis of hemiceluloses and the condensation of lignins. This decreasing in rigidity is certainly the consequence of the hemicellulose degradation. As for the increasing of the softening temperature is surely due to the condensation of the lignins. In fact, the raise of the strong linkages frequency render the polymer network less flexible, and in consequences translate the wood softening to the high temperature. Generally speaking, the new molecular organisation, produced at an elevated temperature level, have a reduced mobility at room temperature.

The response of the fir samples to this experimental protocol is noticeably different. The first exposition at 135°C induces an important reduction of the storage modulus (-25% at 20°C), a raising of the internal friction (the maximum of the loss factor changes from 0.115 to 0.141) and a decrease of the softening temperature of about 4°C. The changing during the following cycles of temperature are tiny. These results suggest an important difference in the biochemical structure of softwood and hardwood. It is possible to formulate the following hypothesis. Heat could break strong inter-molecular bonding in softwood lignins, which are initially highly cross-linked, and by this way could reduce the softening temperature. In order to confirm this assumption, biochemical analysis of treated and untreated samples is scheduled. Moreover, irreversible chemical changes are maybe not the sole explanation for the modifications of viscoelastic properties during cyclic heating/cooling. Actually, Kudo *et al*. [24, 25 ] recently suggested that the modulus of elasticity in bending is fairly reduced by quenching. So, the mechanical properties of wood should be also affected by reversible conformational changes in wood polymers.

## 4. Conclusion

In this paper, the viscoelastic properties of wood under water-saturated conditions have been investigated from 10°C to 135°C using the WAVE$^{T*}$ apparatus.

Dynamic mechanical tests of wood under water-saturated conditions revealed a transition in the vicinity of 70 °C to 105°C depending on the wood species, the wood directions and the frequency. This transition is related to the glass transition of wet *in situ* lignins. In these tests, it is expected that the temperature level solely acts as a mean of activation of the viscoelastic properties. Nevertheless, other tests, performed at constant temperature, proved that, from 80-90°C, the temperature also alters the material itself. The drop of the rigidity of samples could be explained by the degradation of the hemicelluloses. However, above 110-120°C and after several hours of treatment, the tendency is reversed: the storage modulus increases with the treatment duration. In addition, the exposure of wood samples to severe temperature (135°C) for several hours dramatically changes the viscoelastic properties measured at room temperature. This change seems to be consistent with the condensation of lignins as described in the literature for this temperature level.

Dynamic mechanical analysis at high temperature under water-saturated conditions allows the region of thermal activation and thermal degradation to be mapped depending of the wood species and directions. The collected data and parameters could be used to predict and simulate the behaviour of green wood in drying, forming and peeling processes. The unique ability of WAVE$^T$ to assess the mechanical degradation during the treatment could also be applied to define the best parameters for the heat treatment of wood in liquid water (temperature/ time pathway). The present interest to produce ethanol from biomass is certainly a good motivation for this application.

# References

1. Kelley S, Rials T and Glasser W., J. Mater. Sci. 22 (1987) 617-624.
2. Obataya E, Norimoto M and Gril J., Polymer 39(14) (1998) 3059-3064.
3. Salmén L., J. Mater. Sci. 19 (1984) 3090-3096.
4. Olsson A-M. and Salmén L., in "Viscoelasticity of Biomaterials" (A.C. Society, 1992) p. 133-143.
5. Olsson A-M. and Salmén L., in Proceedings of the International Conference of COST Action E8, Mechanical Performance of Wood and Wood Products (1997) 269-279.
6. Gril J., in PhD report, Université Paris 6 (1988).
7. Ranta-Maunus A., Wood Sci. Technol. 9 (1975) 189-205.
8. Ranta-Maunus A., Wood Sci. Technol. 48 (1990) 67-71.
9. Mukudai J. and Yata S., Wood Sci. Technol. 20 (1986) 335-348.
10. Mukudai J. and Yata S., Wood Sci. Technol. 22 (1988) 43-58.
11. Hanhijärvi A. and Hunt D., Wood Sci. Technol. 32 (1998) 57-70.
12. Funaoka M., Kako T. and Abe I., Wood Sci. Technol. 24 (1990) 277-288.
13. Passard J. and Perré P., in Proceedings of the COST E15 Final Conference, Creep of wood at high temperature: thermal activation or thermal degradation? (2004).
14. Takahashi K., Morooka T. and Norimoto M., Wood Res. 85 (1998) 79-80.
15. Hamdan S., Dwianto W., Morooka T. and Norimoto M., Holzforschung 54(5) (2000) 557-560.
16. Passard J. and Perré P., Ann. For. Sci. 62 (2005) 707-716.
17. Passard J. and Perré P., Ann. For. Sci. 62, (2005) 823-830.
18. Placet V, Passard J, Perré P., WAVE$^T$, a custom device able to measure viscoelastic properties of green wood up to 100°C. Submitted to the Maderas Ciença Technologia., November 2007.
19. Placet V., PhD report, University of Nancy 1 (2006).
20. Placet V, Passard J, Perré P., Holzforschung 61-5 (2007) 548-557.
21. Placet V and Perré P., Patent. "Chambre d'essai bi-climatique". Réf: FR0608589 (2006).
22. Timoshenko S.P., "Résistance des matériaux" (Bordas, Paris 1968).
23. Wikberg H and Maunu SL., Polym. Carbohydr. 58 (2004) 461-466.
24. Kudo M., Iida I., Ishimaru Y. and Furuta Y., Mokuzai Gakkaishi. 49 (2003) 253-259.
25. Iida I., Ooi K., Wang Y., Furuta Y.and Ishimaru Y., Mokuzai Gakkaishi. 52 (2006) 93-99.

# Figures

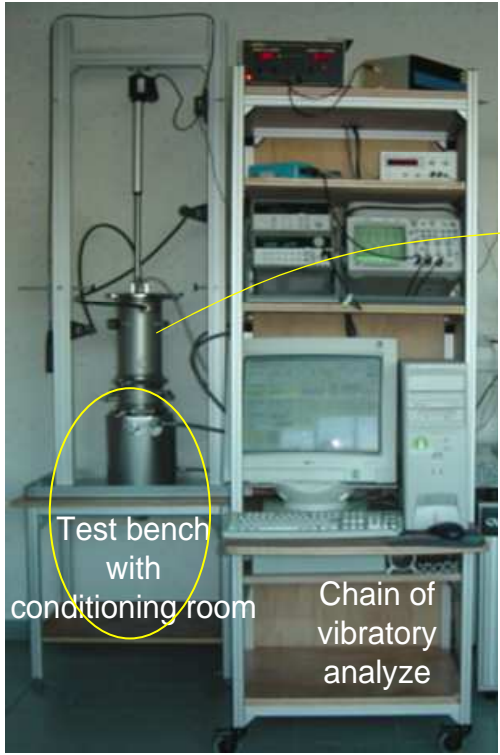 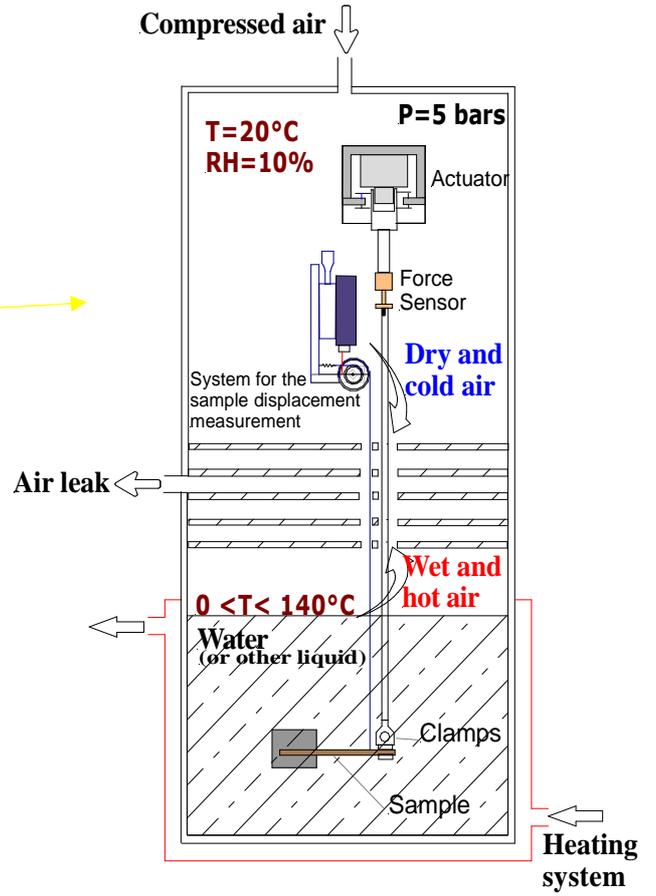

**Figure 1: Experimental apparatus: the WAVE$^T$'s set up**

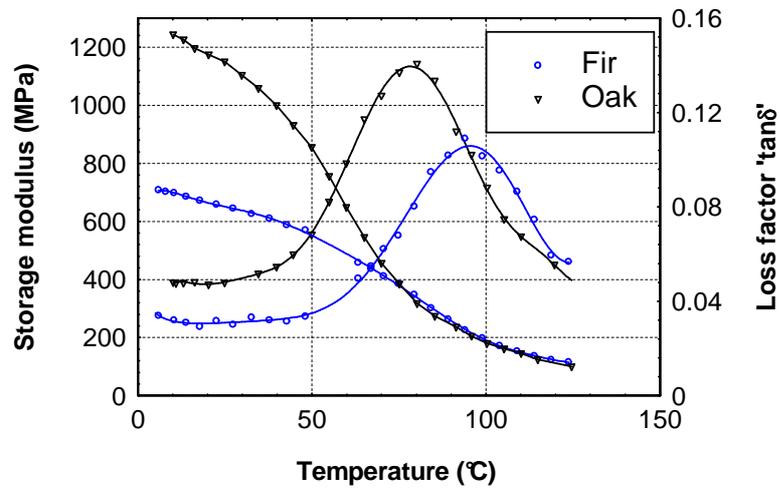

**Figure 2: Thermal activation of the viscoelastic properties of wood in the range 5°C to 125°C. Fir,** *Abies pectinata,* **and Oak,** *Quercus sessiliflora*, **in radial direction, frequency = 1 Hz.**

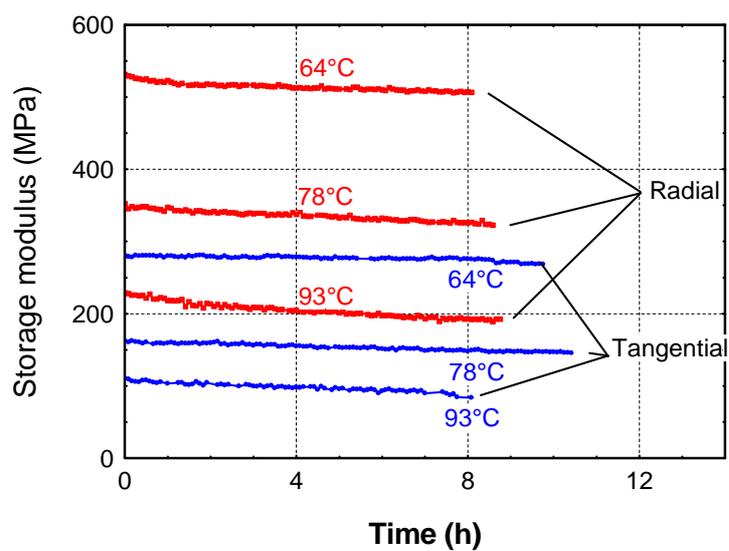

**Figure 3: Evolution of the storage modulus versus time at different plateau temperature (64°C, 78°C, 93°C) for beech samples in radial and tangential directions**

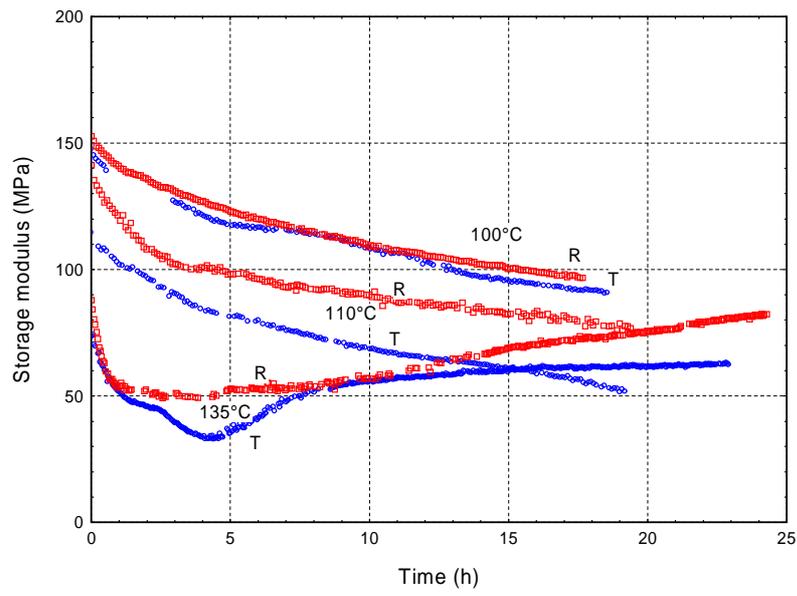

**Figure 4: Evolution of the storage modulus of oak samples in transverse direction according to time at different plateau temperatures (100°C, 110°C and 135°C)**

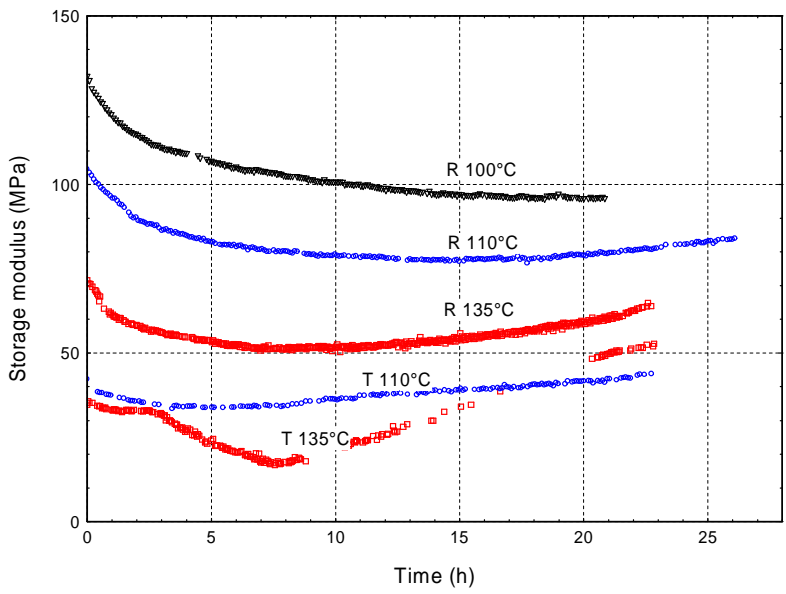

**Figure 5: Evolution of the storage modulus of spruce samples in transverse direction according to time at different plateau temperatures (100°C, 110°C and 135°C)**

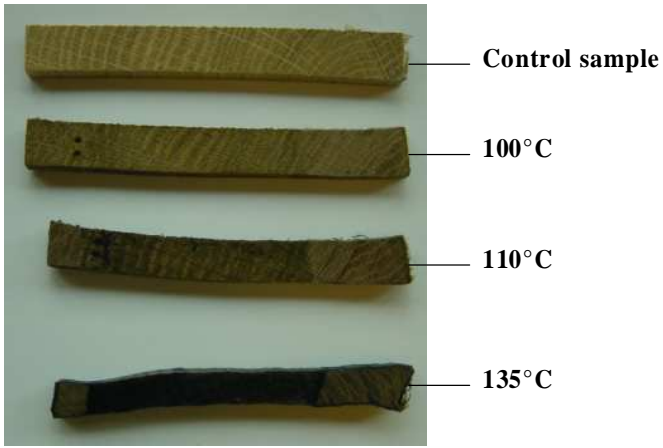

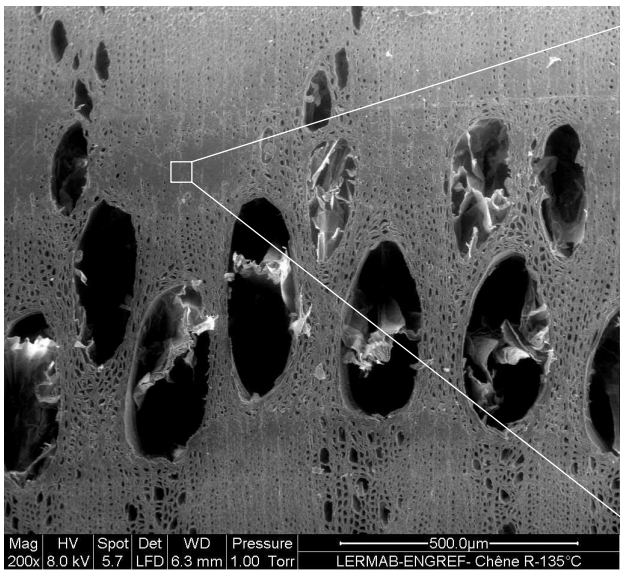
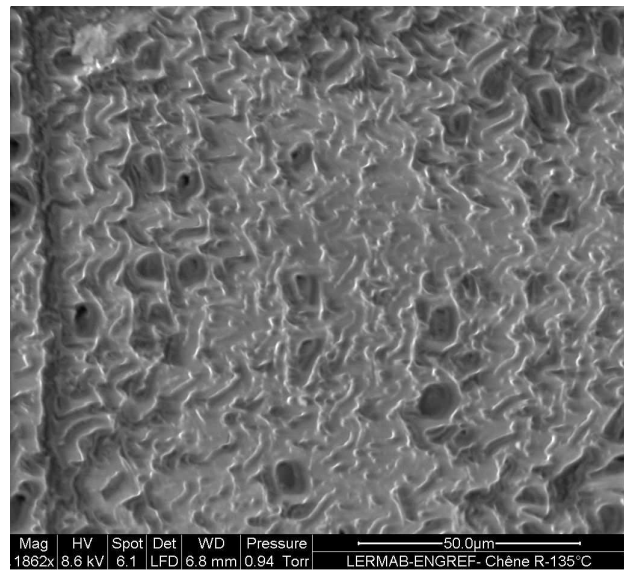

**Figure 6: Anatomical views of treated wood (135°C) at room temperature in moderate conditions (ESEM microphotographs, F.Huber)**

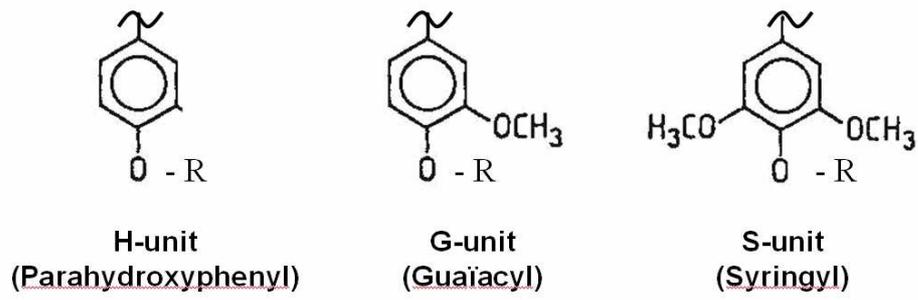

**Figure 7: Constitutive units of Lignin**

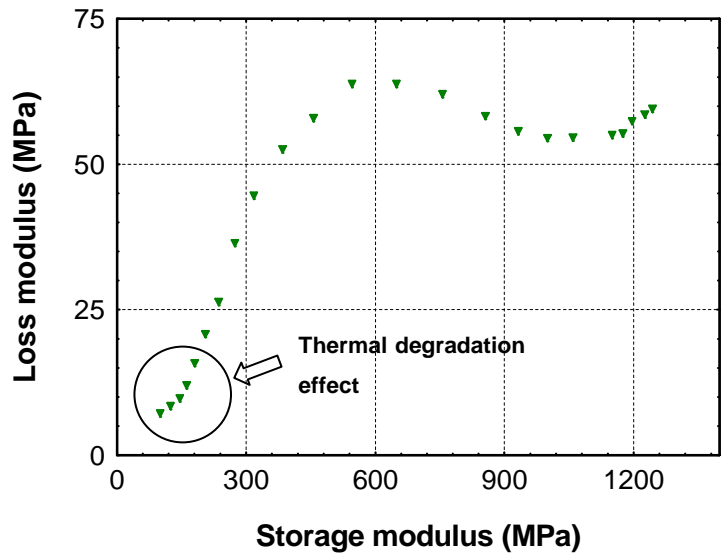

**Figure 8 : Cole-Cole diagram. Data collected for oak sample in radial direction for temperatures between 10°C and 135°C at a frequency of 1 Hz.**

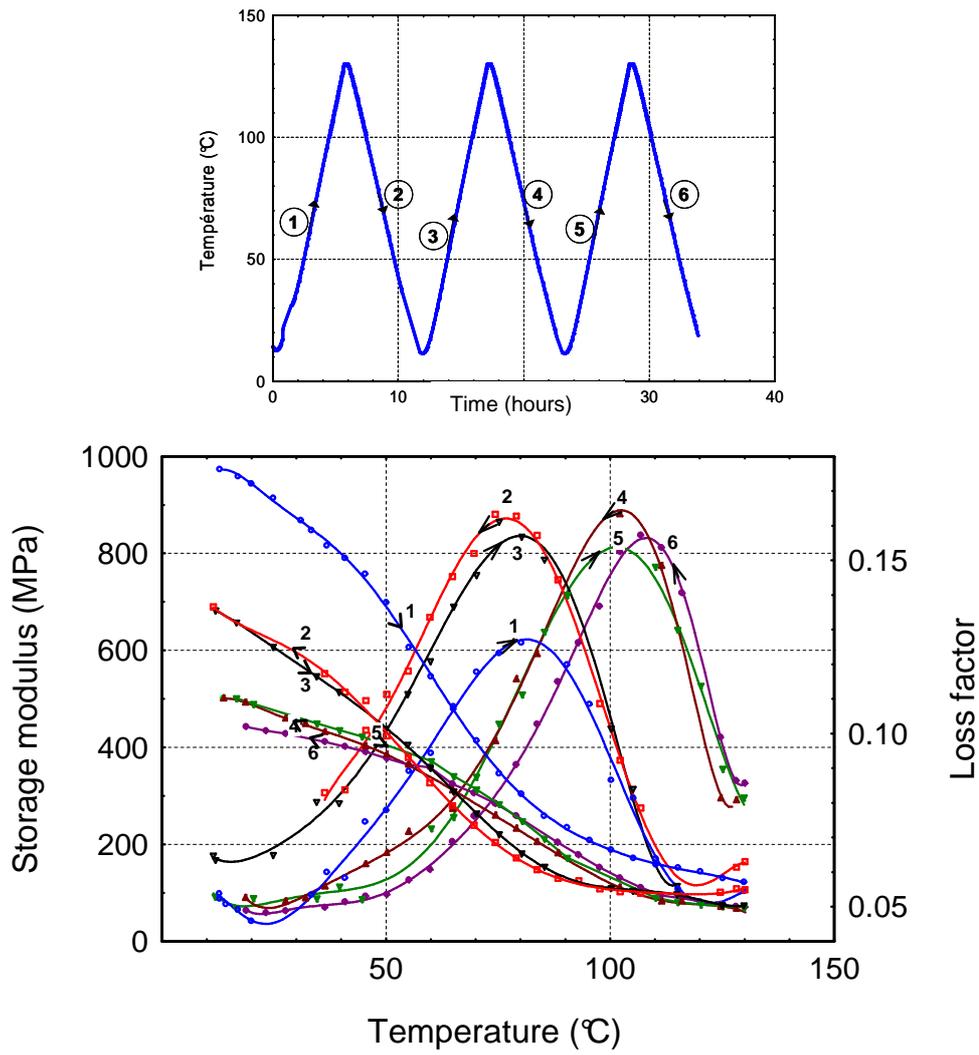

**Figure 9: Evolution of the storage modulus and the loss factor versus temperature for oak submitted to 3 temperature cycles between 10°C and 135°C**

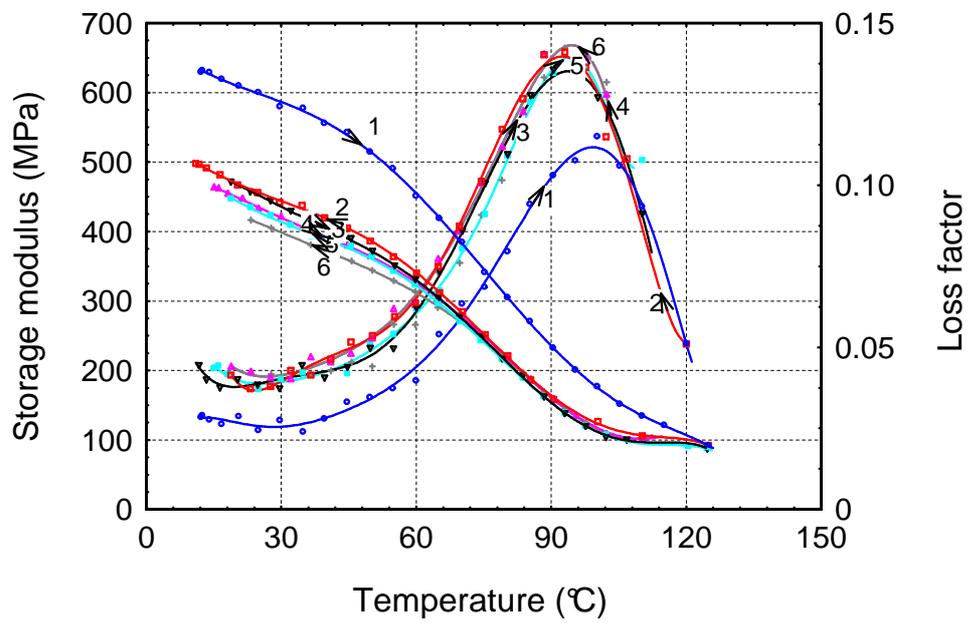

**Figure 10: Evolution of the storage modulus and the loss factor versus temperature for fir submitted to 3 temperature cycles between 10°C and 135°C**

**Tables**

**Table 1: Thermal degradation versus time of beech wood at different temperature plateau (64°C, 78°C, 93°C)**

| T (°C) | | 64°C | | 78°C | | 93°C | |
|---|---|---|---|---|---|---|---|
| Direction | | R | T | R | T | R | T |
| Time (h) | 0 | 530 MPa | 286 MPa | 364 MPa | 174 MPa | 246 MPa | 114 MPa |
| | 2 | - 0.7 % | - 0.7 % | - 1.7 % | - 2.0 % | - 3.7 % | - 4.3 % |
| | 4 | - 1.5 % | - 1.4 % | - 3.3 % | - 4.0 % | - 7.4 % | - 8.6 % |
| | 6 | - 2.2 % | - 2.0 % | - 5.0 % | - 6.0 % | - 11.0 % | - 12.9 % |
| | 8 | - 2.9 % | - 2.7 % | - 6.6 % | - 7.9 % | - 14.7 % | - 17.2 % |

**Table 2: Changes in storage modulus value according to the thermal treatment duration and severity for oak samples.**

| Time (h) | Radial | | | Tangential | | |
|---|---|---|---|---|---|---|
| | 100°C | 110°C | 135°C | 100°C | 110°C | 135°C |
| 0 | 153 MPa | 141 MPa | 88 MPa | 147 MPa | 115 MPa | 73 MPa |
| 0.5 | - 5 % | - 10 % | - 32 % | - 5 % | - 6 % | - 18 % |
| 3.5 | - 16 % | - 29 % | - 44 % | | | |
| 4.5 | | | | - 19 % | - 29 % | - 55 % |
| 8 | - 25 % | - 35 % | - 39 % | - 23 % | - 37 % | - 30 % |
| 14 | - 33 % | - 40 % | - 25 % | - 34 % | - 46 % | - 18 % |
| 18 | - 37 % | - 45 % | - 17 % | - 37 % | - 52 % | - 15 % |
| 20 | | - 46 % | - 15 % | | | - 15 % |
| 24 | | | - 7 % | | | |

**Table 3: Changes in storage modulus value according to the thermal treatment duration and severity for spruce samples.**

| Time (h) | Radial | | | Tangential | |
|---|---|---|---|---|---|
| | 100°C | 110°C | 135°C | 110°C | 135°C |
| 0 | 132 MPa | 104 MPa | 72 MPa | 42 MPa | 35 MPa |
| 0.5 | - 5 % | - 5 % | - 8 % | - 10 % | - 3 % |
| 2 | - 13 % | - 14 % | - 19 % | | |
| 3 | | | | - 19 % | - 9 % |
| 4 | - 17 % | - 19 % | - 25 % | | |
| 8 | - 23 % | - 23 % | - 29 % | - 19 % | - 51 % |
| 12 | - 26 % | - 25 % | - 28 % | - 10 % | - 26 % |
| 16 | - 27 % | - 27 % | - 24 % | - 7 % | 6 % |
| 20 | - 27 % | - 24 % | - 18 % | 0 % | 37 % |
| 24 | | - 21 % | | | |
| 26 | | - 19 % | | | |